\documentclass{article}

\usepackage{arxiv}

\usepackage[utf8]{inputenc}
\usepackage[T1]{fontenc}
\usepackage{lmodern}
\usepackage{graphicx}
\graphicspath{{images/}}
\usepackage{amsmath,amssymb,amsthm}
\usepackage{amsfonts}
\usepackage{booktabs}
\usepackage{multirow}
\usepackage{enumitem}
\usepackage{microtype}
\usepackage{float}
\usepackage{nicefrac}
\usepackage{tabularx}
\usepackage{array}
\usepackage{caption}
\usepackage{subcaption}
\usepackage{xspace}
\usepackage[dvipsnames]{xcolor}
\usepackage{url}
\usepackage{natbib}
\PassOptionsToPackage{colorlinks=true,linkcolor=blue,citecolor=blue,urlcolor=blue}{hyperref}
\usepackage{doi}
\usepackage{hyperref}
\usepackage{cleveref}

\usepackage{tikz}
\usetikzlibrary{shapes.geometric,arrows.meta,positioning,fit,backgrounds,calc}

\usepackage{listings}
\lstdefinestyle{json}{
  language=,
  basicstyle=\ttfamily\small,
  breaklines=true,
  frame=single,
  backgroundcolor=\color{gray!8},
  rulecolor=\color{gray!40},
  numbers=none,
  keepspaces=true,
  showstringspaces=false,
  literate={\ }{\ }1,
}
\lstdefinestyle{pseudo}{
  basicstyle=\ttfamily\small,
  breaklines=true,
  frame=single,
  backgroundcolor=\color{gray!8},
  rulecolor=\color{gray!40},
  numbers=left,
  numberstyle=\tiny\color{gray},
  keepspaces=true,
  showstringspaces=false,
  mathescape=true,
}

\usepackage[ruled,vlined,linesnumbered]{algorithm2e}

\newtheorem{theorem}{Theorem}
\newtheorem{definition}{Definition}

\newcommand{\mcp}{\textsc{MCP}\xspace}
\newcommand{\engine}{\textit{MCP Workflow Engine}\xspace}

\begin{document}
\sloppy

\title{Separating Intelligence from Execution:\\ A Workflow Engine for the Model Context Protocol}

\author{
  Abhinav Singh Parmar \\
  \texttt{abhinav.parmar@infosys.com}
}

\renewcommand{\shorttitle}{Separating Intelligence from Execution}

\hypersetup{
  pdftitle={Separating Intelligence from Execution: A Workflow Engine for the Model Context Protocol},
  pdfsubject={cs.AI, cs.SE},
  pdfauthor={Abhinav Singh Parmar},
  pdfkeywords={Model Context Protocol, LLM Agents, Tool Use, Workflow Orchestration, Token Efficiency},
}

\maketitle

\begin{abstract}
Large Language Model (LLM) agents increasingly interact with external systems through tool-calling protocols such as the Model Context Protocol (\mcp). In prevailing architectures, the agent must reason about \emph{every} tool invocation in \emph{every} session, consuming tokens proportional to the number of actions performed---even when the task has been solved before. We present the \textbf{\mcp Workflow Engine}, a novel \mcp-native orchestration layer that decouples \emph{intelligence} (deciding what to do) from \emph{execution} (carrying it out). An agent reasons \textbf{once} to produce a declarative \emph{workflow blueprint}---a JSON document specifying a directed sequence of \mcp tool calls with parameterized templates, loops, parallel branches, and data piping. Subsequent executions are triggered by a single \texttt{run\_workflow} tool call, consuming one invocation's worth of tokens regardless of the blueprint's internal complexity. We formalize the \textbf{\mcp Mediator} architectural pattern---an \mcp server that simultaneously acts as a client to downstream \mcp servers---and implement it in TypeScript against the \mcp SDK\@. We evaluate the engine on a production-scale Kubernetes CMDB synchronization task spanning \textbf{67 orchestrated steps} across 2 \mcp servers, \textbf{38 namespaces}, \textbf{13 worker nodes}, and \textbf{22 distinct resource types}. The engine reduces per-execution token cost by over \textbf{99\%}, completes the full cluster graph---comprising \textbf{1{,}200+ nodes} and \textbf{2{,}800+ relationships} across \textbf{20 relationship types}---in under 45 seconds, and achieves deterministic, idempotent execution with zero agent involvement at run time.
\end{abstract}


% --- Content ---
\section{Introduction}
\label{sec:intro}

The emergence of tool-augmented LLM agents has transformed software automation. Protocols like the Model Context Protocol (\mcp)~\citep{mcp2024} provide a standardized interface through which agents discover and invoke external tools---from Kubernetes cluster management to graph database queries. Combined with advances in function-calling capabilities~\citep{openai2023function, anthropic2024claude}, agents can now orchestrate complex multi-step operations spanning heterogeneous systems. However, a fundamental inefficiency plagues current agentic architectures: the agent must reason about every action, every time, even for tasks it has solved before.

Consider a recurring infrastructure task: synchronizing all Kubernetes resources into a Configuration Management Database (CMDB) graph~\citep{parmar2026syncscript}. In a production cluster with 38 namespaces and 13 worker nodes, this requires the agent to discover tools across multiple \mcp servers, plan an enumeration strategy for 22 resource types, execute hundreds of API calls with correct relationship semantics, iterate over all namespaces, and handle errors. This process involves 60--100+ tool calls, each requiring the agent to process its growing context window, reason about the next step, and emit a structured invocation. Token consumption grows quadratically as intermediate results accumulate:
\begin{equation}
C_{\text{agent}} = \sum_{i=1}^{N} \bigl(T_{\text{ctx}}^{(i)} + T_{\text{reason}}^{(i)} + T_{\text{call}}^{(i)}\bigr)
\label{eq:agent-cost}
\end{equation}
where $T_{\text{ctx}}^{(i)}$ grows cumulatively with every intermediate API response. The next day, the same task is requested, and the agent pays the full cost again---the intelligence was consumed and discarded.

We observe a clean separation in the lifecycle of such tasks. \textbf{Phase~1---Design} is a one-time, intelligence-intensive phase: the agent explores available tools, learns their schemas through trial invocations, and composes a plan. \textbf{Phase~2---Execution} is the repeated, mechanical phase: tools are called with resolved parameters, results are piped between steps, and errors are retried. This phase requires no intelligence. The key insight is that \emph{intelligence is one-time, but execution is repeated}.

This paper introduces the \engine, a system that crystallizes this separation. The agent's output in Phase~1 is a \textbf{workflow blueprint}---a declarative JSON artifact that can be stored, versioned, shared, and executed without any agent involvement. We formalize the \textbf{\mcp Mediator} pattern---an \mcp server that simultaneously acts as a client to downstream \mcp servers---and provide a reference TypeScript implementation. Evaluated on a production Kubernetes cluster, the engine achieves over 99\% token cost reduction, 40--80$\times$ latency improvement, and deterministic execution of 67 orchestrated steps with zero agent involvement at run time.

We make four contributions:
\begin{enumerate}[leftmargin=*]
  \item The \emph{\mcp Mediator} architectural pattern separating intelligence from execution (\Cref{sec:architecture}).
  \item A minimal five-primitive workflow DSL for \mcp tool orchestration (\Cref{sec:dsl}).
  \item A complete reference implementation in TypeScript (\Cref{sec:implementation}).
  \item A large-scale evaluation demonstrating the system's effectiveness on a production Kubernetes CMDB synchronization task (\Cref{sec:evaluation}).
\end{enumerate}

\section{Related Work}
\label{sec:related}

Our work sits at the intersection of three active research areas: tool-augmented LLM agents, workflow orchestration systems, and multi-agent coordination frameworks. We review each in turn, highlighting the gap that our system addresses.

\paragraph{Tool-Augmented LLM Agents.}
The paradigm of LLMs that interleave reasoning with tool calls was established by ReAct~\citep{yao2023react} and Toolformer~\citep{schick2023toolformer}. ReAct introduced the Observation-Thought-Action loop, where the agent reasons about which tool to invoke, executes it, observes the result, and continues. This loop is powerful but inherently sequential and token-intensive, as each iteration requires the full conversation history to be reprocessed. Subsequent work has extended tool use along several axes: ToolLLM~\citep{qin2023toolllm} curated a benchmark of 16{,}000+ real-world APIs and demonstrated that even frontier models struggle with complex multi-tool compositions; Gorilla~\citep{patil2023gorilla} introduced retrieval-augmented generation for API calls, reducing hallucination of tool names and parameters; and HuggingGPT~\citep{shen2023hugginggpt} showed that an LLM could plan and dispatch tasks across a large model zoo. ART~\citep{paranjape2023art} further decomposed tool use into automatic reasoning chains. These systems have standardized the tool-call interface but not addressed a critical limitation: \emph{the cost of repeated orchestration}. Every execution pays the full reasoning cost, even for identical tasks.

\paragraph{Workflow Orchestration Systems.}
Traditional workflow engines---Apache Airflow~\citep{airflow}, Temporal~\citep{temporal}, Argo Workflows~\citep{argo}, AWS Step Functions~\citep{stepfunctions}, and Prefect~\citep{prefect}---orchestrate multi-step processes as directed acyclic graphs (DAGs). These systems achieve zero LLM cost at execution time because they require no intelligence: every step is pre-specified in code. The taxonomy of workflow management systems~\citep{yu2005taxonomy} established key design dimensions---control flow, data flow, scheduling, and fault tolerance---that inform our DSL design. However, these systems require \emph{developer-authored} DAGs in Python, YAML, or JSON-based state machines. The barrier to creating a new workflow is a software engineering task, not a natural language request. Our contribution bridges this gap: the LLM acts as the workflow designer, producing the DAG artifact automatically, while the engine acts as the executor.

\paragraph{Multi-Agent Frameworks.}
A parallel line of work has developed frameworks for building LLM agents with tool access: LangChain~\citep{langchain}, CrewAI~\citep{crewai}, AutoGen~\citep{wu2023autogen}, MetaGPT~\citep{hong2023metagpt}, and ChatDev~\citep{qian2023chatdev}. These systems provide powerful abstractions for role-playing agents, multi-agent conversations, and compositional chains. However, they operate at the \emph{agent level}: the LLM remains in the loop for every step of execution. Even when chains are predefined, each step typically involves an LLM call for reasoning, routing, or formatting. Our system operates at the \emph{infrastructure level}: after design time, the LLM is entirely removed from the execution loop. The distinction is analogous to the difference between an interpreter and a compiler---agent frameworks interpret each step through the LLM, whereas our engine compiles the agent's plan into an executable artifact that runs without further interpretation.

\paragraph{Planning and Cost Optimization.}
Recent work on LLM planning~\citep{hao2023reasoning, wang2023plan} has explored how agents can decompose complex tasks into subgoals, and cost optimization work such as FrugalGPT~\citep{chen2023frugalgpt} has focused on reducing inference costs through model cascading and caching. Comprehensive surveys of LLM-based agents~\citep{wang2024survey, xi2023rise} identify tool use, planning, and memory as core capabilities, but do not examine the repeated-cost problem that arises when the same multi-step task is executed multiple times. Code-first agent frameworks like TaskWeaver~\citep{qiao2023taskweaver} and SWE-agent~\citep{yang2024swe} generate executable code rather than declarative plans, offering expressiveness at the cost of sandboxing and inspectability.

\paragraph{Positioning of Our Work.}
\Cref{tab:comparison} summarizes the positioning of our system relative to existing approaches. The critical insight is that ReAct-style systems pay LLM token costs on \emph{every} execution, while traditional workflow engines achieve zero execution cost but require human developers to author DAGs. Our system uses agent tokens \emph{once}---to explore available \mcp tools, understand their schemas, and compose a workflow blueprint---after which execution proceeds with zero LLM involvement.

\begin{table}[H]
\centering
\caption{Comparison of orchestration approaches for multi-step tool-using tasks. We distinguish \emph{design cost} (one-time) from \emph{execution cost} (per-run). $N$ = number of steps, $\bar{T}$ = average tokens per step.}
\label{tab:comparison}
\small
\resizebox{\textwidth}{!}{%
\begin{tabular}{lcccc}
\toprule
\textbf{System} & \textbf{LLM in exec.} & \textbf{Reusable artifact} & \textbf{\mcp-native} & \textbf{Exec.\ cost/run} \\
\midrule
ReAct/Toolformer~\citep{yao2023react,schick2023toolformer}
  & Yes & None & No & $O(N \cdot \bar{T})$ \\
LangChain/CrewAI~\citep{langchain,crewai}
  & Yes & Chains (code) & No & $O(N \cdot \bar{T})$ \\
AutoGen/MetaGPT~\citep{wu2023autogen,hong2023metagpt}
  & Yes & Conversations & No & $O(N \cdot \bar{T})$ \\
Airflow/Temporal~\citep{airflow,temporal}
  & No & DAGs (code) & No & 0 (no LLM) \\
TaskWeaver~\citep{qiao2023taskweaver}
  & Yes & Code & No & $O(N \cdot \bar{T})$ \\
\textbf{\mcp Workflow Engine}
  & \textbf{No} & \textbf{JSON blueprints} & \textbf{Yes} & $\mathbf{O(1)}$ \\
\bottomrule
\end{tabular}%
}
\end{table}

\section{System Design}
\label{sec:design}

\subsection{Architecture: The MCP Mediator Pattern}
\label{sec:architecture}

The \engine occupies a unique architectural position: it is simultaneously an \textbf{\mcp server} (exposing workflow management tools to agents) and an \textbf{\mcp client} (connecting to downstream \mcp servers to execute tool calls). We term this the \textbf{\mcp Mediator} pattern, drawing an analogy to the Mediator design pattern~\citep{gof1994} in which a central object coordinates interactions between decoupled components.

\begin{figure}[H]
\centering
\includegraphics[width=0.95\textwidth]{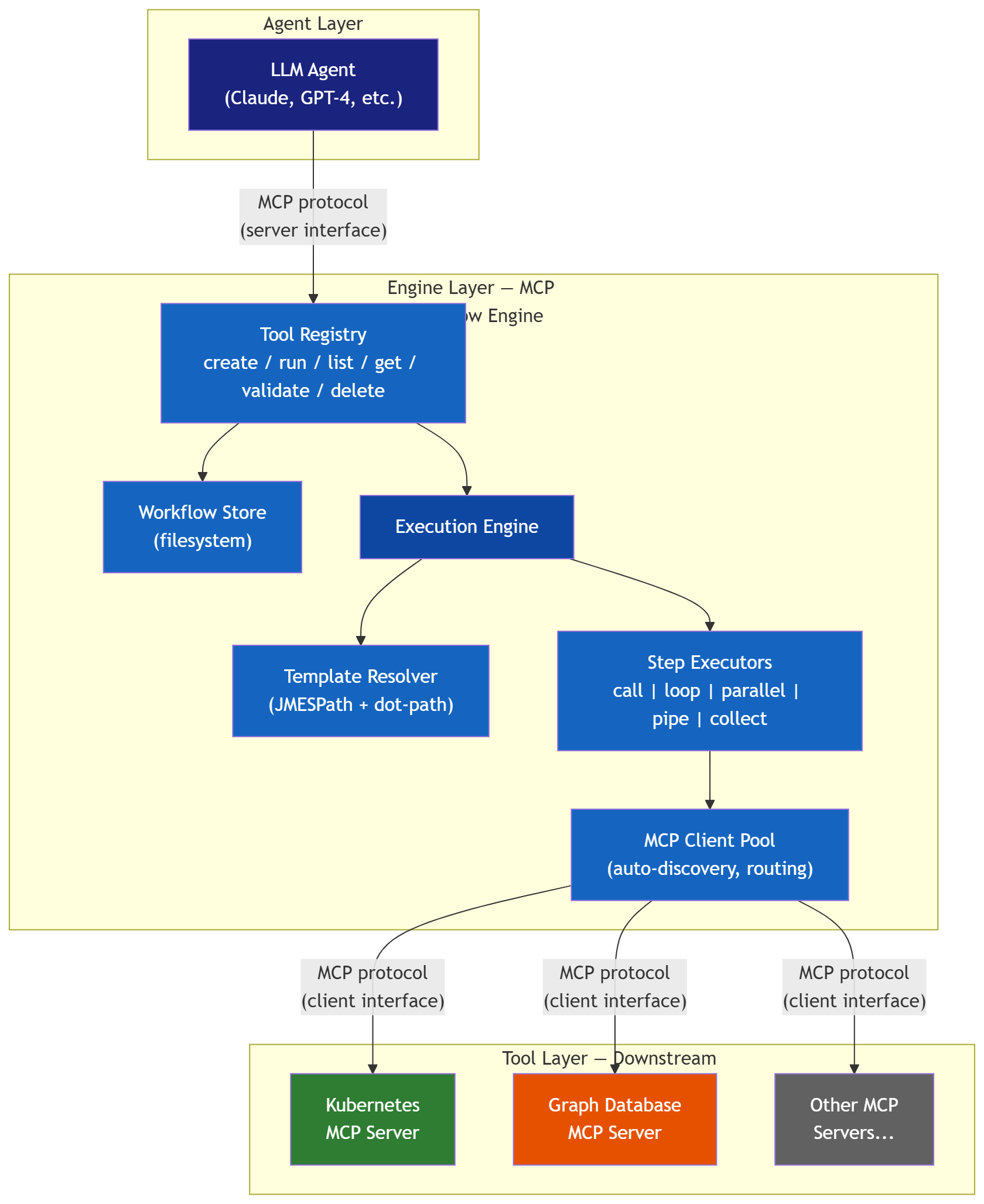}
\caption{Three-layer architecture of the \mcp Workflow Engine (MCP Mediator pattern). The engine is simultaneously an \mcp server (upward, exposing workflow tools to agents) and an \mcp client (downward, connecting to Kubernetes, Graph DB, and other downstream \mcp servers).}
\label{fig:architecture}
\end{figure}

\Cref{fig:architecture} shows the three-layer architecture. The engine bridges the agent layer and the tool layer, removing the agent from the execution path during Phase~2. The Mediator pattern yields three key properties: (1)~\emph{Transparency}---downstream \mcp servers require no modification and are unaware of the engine's existence; (2)~\emph{Composability}---any \mcp-compliant tool server is automatically available to workflows via auto-discovery; and (3)~\emph{Separation of concerns}---the agent interacts only with workflow-level abstractions (\texttt{create\_workflow}, \texttt{run\_workflow}) while the engine handles connection management, routing, retries, and data flow.

\subsection{Two-Phase Lifecycle}
\label{sec:lifecycle}

\begin{figure}[H]
\centering
\includegraphics[width=0.95\textwidth]{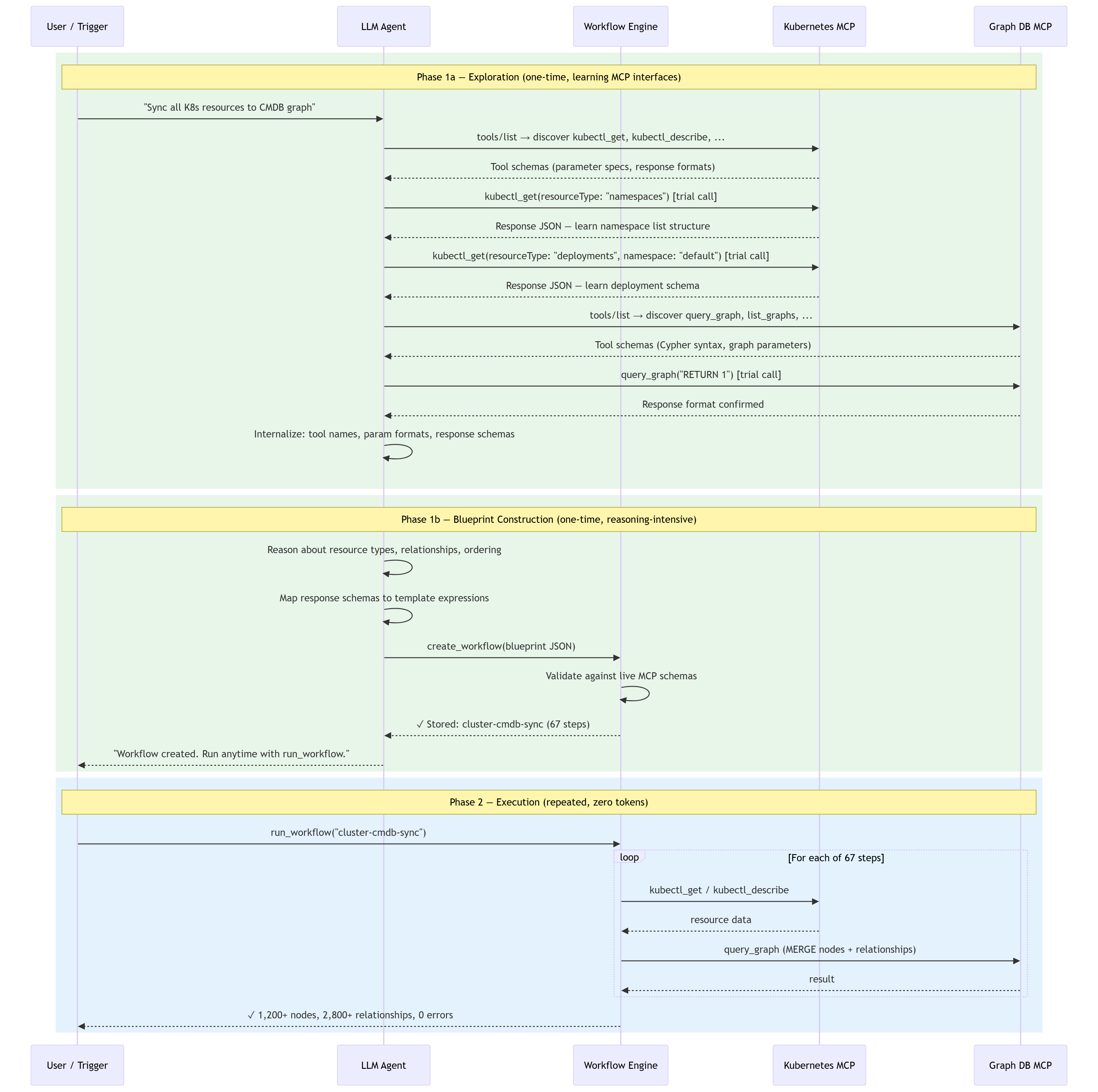}
\caption{Two-phase lifecycle. The LLM agent is involved only at design time (Phase~1). All subsequent executions are handled entirely by the Workflow Engine (Phase~2), consuming zero agent tokens.}
\label{fig:lifecycle}
\end{figure}

\Cref{fig:lifecycle} illustrates the system's two-phase lifecycle. During \textbf{Phase~1a (Exploration)}, the agent discovers downstream \mcp servers by calling \texttt{tools/list} to enumerate available tools and their JSON-Schema parameter specifications, and makes trial invocations to observe real response structures. During \textbf{Phase~1b (Blueprint Construction)}, the agent reasons about resource types, relationships, and ordering to produce a declarative workflow blueprint. Both sub-phases are token-intensive but happen once. During \textbf{Phase~2 (Execution)}, the stored blueprint is triggered by a single \texttt{run\_workflow} call, and the engine autonomously executes all steps with zero agent involvement.

\subsection{The Workflow Blueprint}
\label{sec:blueprint}

A workflow blueprint is the \textbf{reusable artifact}---the output of the agent's intelligence. It is a JSON document specifying an identifier, description, parameterized inputs, an error handling strategy, and an ordered sequence of steps. The blueprint is \emph{declarative} (describes what to do, not how), \emph{parameterized} (runtime values are injected via \texttt{\{\{params.*\}\}} templates), \emph{portable} (JSON files that can be versioned in Git, shared across teams, and scheduled via cron), and \emph{inspectable} (human- and machine-readable, enabling validation and audit).

\begin{lstlisting}[style=json, caption={Workflow blueprint schema (abridged).}, label={lst:blueprint}]
{
  "id": "cluster-cmdb-sync",
  "description": "Sync full Kubernetes cluster into CMDB graph",
  "version": "2.0.0",
  "params": {
    "graph":   { "type": "string", "default": "cmdb-prod" },
    "cluster": { "type": "string", "default": "atlas-prod" }
  },
  "errorStrategy": {
    "onStepFailure": "continue",
    "maxRetries": 2, "retryDelayMs": 1000
  },
  "steps": [ /* Step[] -- see Section 3.4 */ ]
}
\end{lstlisting}

\subsection{The Step DSL}
\label{sec:dsl}

We deliberately constrain the DSL to \textbf{five step types}, sufficient for the vast majority of multi-tool orchestration patterns while keeping the system's semantics formal and its failure modes predictable. This constraint is intentional: JSON-based DSLs tend to evolve into accidental programming languages, and a firm boundary avoids that fate.

\begin{enumerate}[leftmargin=*]
\item \textbf{\texttt{call}}---Invoke any \mcp tool. The atomic unit of execution; calls a named tool on a downstream \mcp server with template-resolved parameters.

\item \textbf{\texttt{loop}}---Iterate over a collection. Executes a sub-step for each item in a resolved array, injecting per-item context (e.g., iterating over namespaces).

\item \textbf{\texttt{parallel}}---Concurrent execution. Runs multiple independent branches concurrently via \texttt{Promise.allSettled}, maximizing throughput for independent operations.

\item \textbf{\texttt{pipe}}---Sequential chain. Each step receives the output of the previous, enabling data transformation pipelines.

\item \textbf{\texttt{collect}}---Batch results. Aggregates outputs from a sub-workflow for hybrid agent-engine patterns, where the agent reviews batch results rather than individual items.
\end{enumerate}

\begin{figure}[H]
\centering
\includegraphics[width=0.85\textwidth]{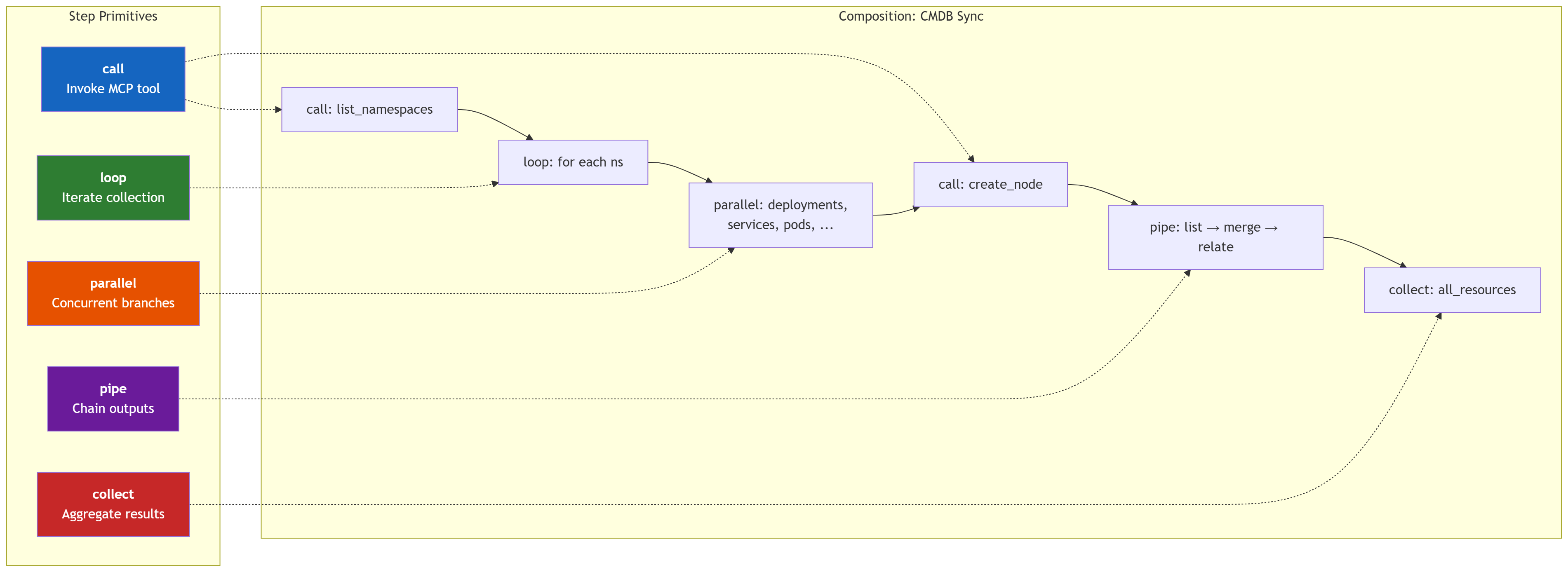}
\caption{The five step primitives (\texttt{call}, \texttt{loop}, \texttt{parallel}, \texttt{pipe}, \texttt{collect}) and their composition in the CMDB synchronization workflow.}
\label{fig:primitives}
\end{figure}

\Cref{fig:primitives} illustrates the five step primitives and their composition in a concrete CMDB sync example.

We intentionally exclude constructs common in full-featured workflow engines: \emph{conditionals} (branching requires the kind of reasoning best left to the agent), \emph{variables} (data flows implicitly through the execution context via \texttt{steps.<id>}, eliminating scoping bugs), and \emph{string manipulation} (data shaping remains the agent's responsibility). The design rationale is discussed further in \Cref{sec:discussion-simplicity}.

\subsection{Template Resolution}
\label{sec:template}

\begin{figure}[H]
\centering
\includegraphics[width=0.85\textwidth]{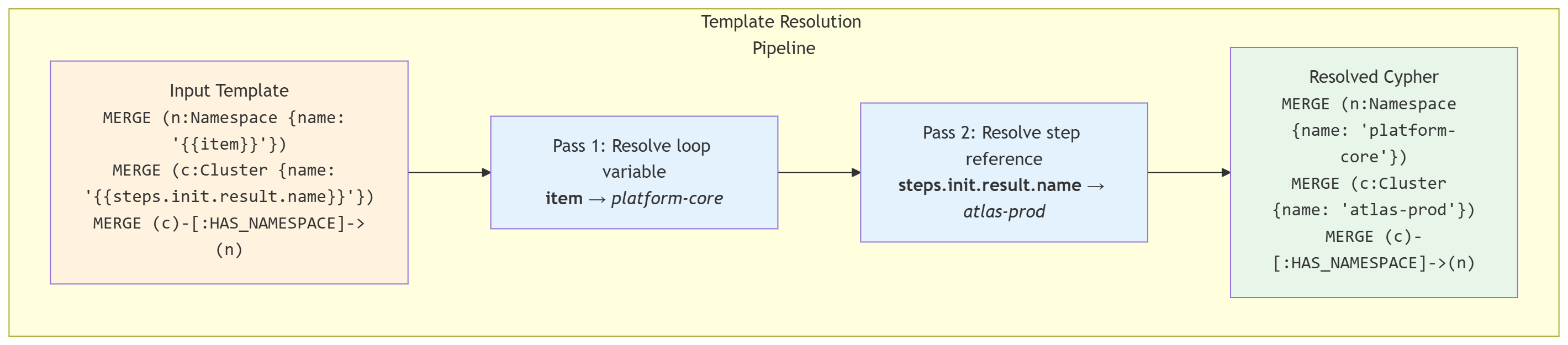}
\caption{Template resolution pipeline. The resolver performs multi-pass substitution, resolving loop variables and step references to produce a fully-resolved Cypher query.}
\label{fig:template}
\end{figure}

Parameters in step definitions use \texttt{\{\{expression\}\}} syntax. The resolver supports two strategies: (1)~JMESPath expressions~\citep{jmespath} for structured data extraction, and (2)~dot-path fallback for simple property access. A critical design decision: when the entire value is a single template expression, the resolver returns the \textbf{raw value} (preserving arrays and objects) rather than stringifying, enabling type-safe data flow between steps. \Cref{fig:template} illustrates the multi-pass template resolution pipeline.

\subsection{Execution Context and Algorithm}
\label{sec:algorithm}

The engine maintains a \texttt{ResolverContext} that accumulates state as steps execute. Each completed step's output is stored at \texttt{steps.<stepId>}, making it available to all subsequent steps via template expressions. This provides implicit data flow without explicit variable declarations. \Cref{alg:execute} presents the execution pseudocode.

\begin{algorithm}[H]
\DontPrintSemicolon
\caption{Workflow Execution}
\label{alg:execute}
\KwIn{Blueprint $B$, Parameters $P$}
\KwOut{Execution result $R$}

$\mathit{ctx} \leftarrow \{\,\text{params}: P,\; \text{steps}: \{\}\,\}$\;
$\mathit{results} \leftarrow [\;]$\;

\ForEach{step $s$ in $B.\text{steps}$}{
  $\mathit{rp} \leftarrow \textsc{ResolveTemplates}(s.\text{params}, \mathit{ctx})$\;
  \Switch{$s.\text{type}$}{
    \Case{\texttt{call}}{
      $\mathit{result} \leftarrow \textsc{ClientPool.CallTool}(s.\text{tool}, \mathit{rp})$\;
    }
    \Case{\texttt{loop}}{
      $\mathit{items} \leftarrow \textsc{Resolve}(s.\text{over}, \mathit{ctx})$\;
      \ForEach{item in items}{
        $\mathit{ctx}[s.\text{as}] \leftarrow \mathit{item}$\;
        $\mathit{result}.\text{append}(\textsc{ExecuteStep}(s.\text{do}, \mathit{ctx}))$\;
      }
    }
    \Case{\texttt{parallel}}{
      $\mathit{result} \leftarrow \textsc{AllSettled}(s.\text{branches}.\text{map}(\textsc{ExecStep}))$\;
    }
    \Case{\texttt{pipe}}{$\mathit{result} \leftarrow \textsc{Chain}(s.\text{steps}, \mathit{ctx})$\;}
    \Case{\texttt{collect}}{$\mathit{result} \leftarrow \textsc{Aggregate}(s, \mathit{ctx})$\;}
  }
  $\mathit{ctx}.\text{steps}[s.\text{id}] \leftarrow \mathit{result}$\;
  $\mathit{results}.\text{append}(\{\,\text{stepId}: s.\text{id},\; \text{result}\,\})$\;
  \If{$\mathit{result}.\text{error}$ \textbf{and} $B.\text{errorStrategy} = \texttt{abort}$}{
    \Return $\textsc{Failure}(\mathit{results})$\;
  }
}
\Return $\textsc{Success}(\mathit{results})$\;
\end{algorithm}

\subsection{MCP Client Pool and Error Handling}
\label{sec:error}

The client pool bridges the engine to downstream \mcp servers. On initialization, it connects to all configured servers (stdio, SSE, or streamable-HTTP transport), discovers tools by calling \texttt{tools/list} on each server, and builds a routing table $\text{toolName} \to \text{serverClient}$. Every tool is treated as a pure function: $(\text{name}, \text{params}) \to \text{result}$.

The engine supports three error strategies (\Cref{tab:error}): \texttt{abort} (stop on first failure), \texttt{continue} (log and proceed), and \texttt{retry} (retry with backoff). When \texttt{collectErrors: true}, all errors are accumulated alongside successful results, enabling partial-success workflows essential for large-scale infrastructure operations. The engine also performs two-tier validation at design time: structural errors (missing IDs, unknown types) block saving, while tool warnings (tools not found in connected servers) produce non-blocking warnings.

\begin{table}[H]
\centering
\caption{Error handling strategies.}
\label{tab:error}
\small
\begin{tabular}{lll}
\toprule
\textbf{Strategy} & \textbf{Behavior} & \textbf{Use Case} \\
\midrule
\texttt{abort}    & Stop immediately on first failure   & Critical pipelines \\
\texttt{continue} & Log error, skip step, proceed       & Best-effort sync \\
\texttt{retry}    & Retry up to $n$ times with backoff  & Transient failures \\
\bottomrule
\end{tabular}
\end{table}

\section{Token Cost Analysis}
\label{sec:cost}

\subsection{Formal Cost Model}

We model the token cost of a multi-step tool-using task under two architectures.

\begin{definition}[Agent-in-the-loop cost]
For a task requiring $N$ tool calls, the total token cost under the baseline agent architecture is:
\begin{equation}
C_{\text{agent}} = T_{\text{task}} + \sum_{i=1}^{N} \left(
  \underbrace{\sum_{j=1}^{i-1} T_{\text{result}}^{(j)}}_{\text{accumulated context}}
  + T_{\text{reason}}^{(i)} + T_{\text{call}}^{(i)} \right)
\end{equation}
The accumulated context term gives worst-case complexity $O(N^2 \cdot \bar{T}_{\text{result}})$.
\end{definition}

\begin{definition}[Workflow engine cost]
Under the proposed architecture:
\begin{align}
C_{\text{design}} &= T_{\text{task}} + T_{\text{explore}} + T_{\text{reasoning}} + T_{\text{blueprint}} \\
C_{\text{exec}}   &= T_{\text{trigger}} + T_{\text{call}} \approx 150\ \text{tokens}
\end{align}
where $T_{\text{explore}}$ captures the agent's exploratory interaction with downstream \mcp servers (discovering tools, making trial calls, learning response schemas).
\end{definition}

\begin{theorem}[Amortized cost]
Over $K$ executions: $C_{\text{agent}}^{(K)} = K \cdot C_{\text{agent}}$ and $C_{\text{engine}}^{(K)} = C_{\text{design}} + K \cdot C_{\text{exec}}$.
Break-even occurs at $K^{*} = C_{\text{design}} / (C_{\text{agent}} - C_{\text{exec}}) \approx 0.04$, meaning the engine recovers its design cost after approximately 4\% of one full agent execution.
\end{theorem}

\subsection{Empirical Measurement}

We measure token costs on the CMDB synchronization task: 67 steps across 2 \mcp servers, syncing 22 resource types from a Kubernetes cluster with 38 namespaces and 13 nodes into a graph database. Detailed per-row derivations follow the unit cost assumption of ${\sim}500$ input tokens and ${\sim}100$ output tokens per agent step (see \Cref{sec:appendix-tokens}).

\begin{table}[H]
\centering
\caption{Estimated token consumption: Agent-in-the-loop baseline.}
\label{tab:agent-tokens}
\small
\begin{tabular}{lrrrr}
\toprule
\textbf{Phase} & \textbf{Steps} & \textbf{Input} & \textbf{Output} & \textbf{Total} \\
\midrule
Tool discovery \& planning               &    3        & 1{,}500      & 2{,}500     & 4{,}000     \\
Cluster-scoped fetches                   &    5        & 5{,}000      & 2{,}500     & 7{,}500     \\
Namespace enumeration                    &    1        & 500          & 1{,}200     & 1{,}700     \\
Per-ns resource fetches (38$\times$12)   &  456        & 228{,}000    & 45{,}600    & 273{,}600   \\
Graph node creation loops                & $\sim$800   & 400{,}000    & 80{,}000    & 480{,}000   \\
Relationship creation (20$\times$38)     & $\sim$760   & 380{,}000    & 76{,}000    & 456{,}000   \\
Error recovery \& re-planning            & $\sim$15    & 15{,}000     & 3{,}000     & 18{,}000    \\
Summary synthesis                        &    1        & 5{,}000      & 1{,}000     & 6{,}000     \\
\midrule
\textbf{Total per execution}             & $\mathbf{\sim\!2{,}041}$ & $\mathbf{\sim\!1{,}035{,}000}$ & $\mathbf{\sim\!211{,}800}$ & $\mathbf{\sim\!1{,}246{,}800}$ \\
\bottomrule
\end{tabular}
\end{table}

\begin{table}[H]
\centering
\caption{Token consumption: Workflow engine.}
\label{tab:engine-tokens}
\small
\begin{tabular}{lrrrr}
\toprule
\textbf{Phase} & \textbf{Cycles} & \textbf{Input} & \textbf{Output} & \textbf{Total} \\
\midrule
Exploration (one-time)              & $\sim$8 tool-calling & $\sim$12{,}000 & $\sim$4{,}000  & $\sim$16{,}000 \\
Blueprint construction (one-time)   & $\sim$15 reasoning   & $\sim$30{,}000 & $\sim$8{,}000  & $\sim$38{,}000 \\
\midrule
\textbf{Total design (one-time)}    & $\sim$23             & $\sim$42{,}000 & $\sim$12{,}000 & $\sim$54{,}000 \\
\midrule
Execute (per run)                   & 1 tool call          & $\sim$100      & $\sim$50       & $\sim$150      \\
\bottomrule
\end{tabular}
\end{table}

\begin{table}[H]
\centering
\caption{Amortized token cost comparison over $K$ executions. The engine's cost is $54{,}000 + K \times 150$.}
\label{tab:amortized}
\small
\begin{tabular}{rrrl}
\toprule
\textbf{Runs ($K$)} & \textbf{Agent Total} & \textbf{Engine Total} & \textbf{Token Savings} \\
\midrule
1   & 1{,}246{,}800     & 54{,}150  & 95.7\%    \\
2   & 2{,}493{,}600     & 54{,}300  & 97.8\%    \\
5   & 6{,}234{,}000     & 54{,}750  & 99.1\%    \\
10  & 12{,}468{,}000    & 55{,}500  & 99.6\%    \\
50  & 62{,}340{,}000    & 61{,}500  & 99.9\%    \\
365 & 455{,}082{,}000   & 108{,}750 & $\sim$99.98\%   \\
\bottomrule
\end{tabular}
\end{table}

The per-row figures in \Cref{tab:agent-tokens} use a flat 500-token-per-step approximation as a conservative lower bound, deliberately omitting the cumulative context growth term from \Cref{eq:agent-cost}. Real agent costs are higher; all reported savings percentages are therefore \emph{understated}.

\subsection{Beyond Tokens: Latency and Reliability}

Token cost is the primary metric, but two additional dimensions merit discussion. \textbf{Latency:} Each agent reasoning cycle introduces 1--5~seconds of inference latency. A 67-step workflow with hundreds of loop iterations would require 30--60 minutes of wall-clock time under agent-in-the-loop execution. The workflow engine completes the same task in \textbf{under 45 seconds}---a 40--80$\times$ improvement bounded only by downstream API response times. \textbf{Reliability:} Agent-in-the-loop execution is \emph{probabilistic}---the agent may hallucinate tool names, construct invalid queries, or exceed the context window mid-task. The workflow engine's execution is \emph{deterministic}: the same blueprint with the same cluster state produces the same tool calls every time. Idempotent \texttt{MERGE} semantics ensure convergent state even across retries.

\section{Evaluation}
\label{sec:evaluation}

\subsection{Experimental Setup}

We evaluate the system on a production Kubernetes cluster\footnote{All cluster names, namespace names, node identifiers, and resource names have been anonymized. See \Cref{sec:appendix-anon}.} whose characteristics are summarized in \Cref{tab:cluster}.

\begin{table}[H]
\centering
\caption{Target cluster characteristics.}
\label{tab:cluster}
\small
\begin{tabular}{ll}
\toprule
\textbf{Property} & \textbf{Value} \\
\midrule
Cluster name       & \textit{atlas-prod} (anonymized) \\
Worker nodes       & 13 (\texttt{node-alpha-01} through \texttt{node-alpha-13}) \\
Namespaces         & 38 (platform-core, data-services, ml-workloads, \ldots) \\
Resource types synced & 22 \\
Downstream \mcp servers & 2 (Kubernetes, Graph Database) \\
Graph database     & Property graph (Cypher query language) \\
\bottomrule
\end{tabular}
\end{table}

\subsection{Full-Cluster CMDB Synchronization}

\begin{figure}[H]
\centering
\includegraphics[width=0.95\textwidth]{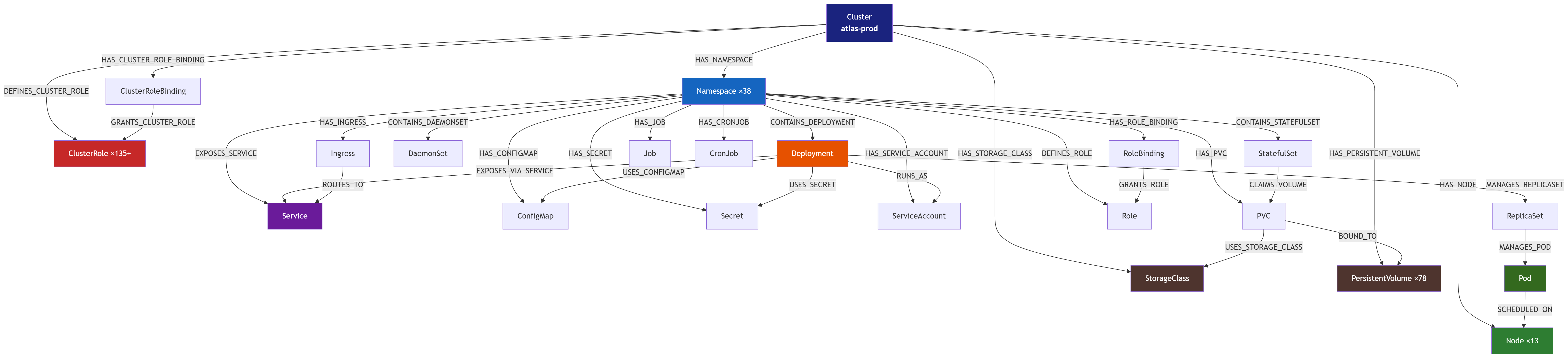}
\caption{CMDB graph schema produced by the workflow, showing 16 node labels and 20 relationship types. The Cluster node is the root, with Namespace and Node as primary children.}
\label{fig:graph-schema}
\end{figure}

The \texttt{cluster-cmdb-sync} workflow blueprint contains 67 top-level steps organized into three phases (\Cref{tab:workflow-structure}): cluster-scoped resource creation (8 steps), namespace-scoped resource iteration (39 steps using \texttt{call} and \texttt{loop}), and relationship construction across 20 relationship types (20 steps). \Cref{fig:graph-schema} illustrates the resulting CMDB graph schema.

\begin{table}[H]
\centering
\caption{Workflow structure by phase.}
\label{tab:workflow-structure}
\small
\begin{tabular}{llrl}
\toprule
\textbf{Phase} & \textbf{Description} & \textbf{Steps} & \textbf{Types} \\
\midrule
1.\ Cluster-scoped  & Cluster node, 13 nodes, storage, PVs, roles, CRBs & 8  & \texttt{call} \\
2.\ Namespace-scoped & 38 ns $\times$ 15 resource types & 39 & \texttt{call}, \texttt{loop} \\
3.\ Relationships    & 20 types linking resources & 20 & \texttt{call} \\
\midrule
\textbf{Total}       & & \textbf{67} & \\
\bottomrule
\end{tabular}
\end{table}

\Cref{tab:execution} presents the execution metrics. The engine completed all 67 top-level steps (expanding to $\sim$2{,}000+ total \mcp tool invocations via loop iterations), producing a graph of 1{,}200+ nodes and 2{,}800+ relationships with zero errors and zero agent tokens.

\begin{table}[H]
\centering
\caption{Full-cluster CMDB sync execution metrics.}
\label{tab:execution}
\small
\begin{tabular}{ll}
\toprule
\textbf{Metric} & \textbf{Value} \\
\midrule
Total execution time                              & $\sim$42 seconds \\
Top-level steps executed                          & 67 \\
Total \mcp tool invocations (incl.\ loop iters)   & $\sim$2{,}000+ \\
Graph nodes created                               & 1{,}200+ \\
Graph relationships created                       & 2{,}800+ \\
Distinct relationship types                       & 20 \\
Distinct node labels                              & 16 \\
Errors (with \texttt{continue} strategy)          & 0 \\
Agent tokens consumed (execution)                 & 0 \\
\bottomrule
\end{tabular}
\end{table}

\subsection{Scalability}

To demonstrate scalability, we compare single-namespace and full-cluster configurations (\Cref{tab:scaling}). The key observation: token cost does not scale with cluster size under the workflow engine, whereas under the agent baseline it grows super-linearly due to context accumulation.

\begin{table}[H]
\centering
\caption{Scaling from single-namespace to full-cluster sync.}
\label{tab:scaling}
\small
\begin{tabular}{lrr}
\toprule
\textbf{Metric} & \textbf{Single-Namespace} & \textbf{Full-Cluster} \\
\midrule
Namespaces          & 1         & 38          \\
Blueprint steps     & 27        & 67          \\
Resource types      & 8         & 22          \\
Loop iterations     & $\sim$138 & $\sim$2{,}000+ \\
Graph nodes         & 139       & 1{,}200+    \\
Relationships       & 245       & 2{,}800+    \\
Execution time      & $\sim$20s & $\sim$42s   \\
Agent tokens/run    & 0         & 0           \\
\bottomrule
\end{tabular}
\end{table}

\subsection{Parameterized Re-execution and Idempotency}

The same workflow can be re-executed with different runtime parameters (e.g., \texttt{run\_workflow("cluster-cmdb-sync", \{"graph": "cmdb-staging"\})}) to target a separate graph instance, demonstrating parameterization without blueprint modification. The workflow exclusively uses Cypher \texttt{MERGE} operations, making it idempotent---re-running updates timestamps and adds resources without creating duplicates, essential for scheduled hourly synchronization.

\subsection{Component Ablation}

\Cref{tab:ablation} presents ablation results. Template resolution is essential (all steps fail without it); parallel execution reduces execution time by $\sim$38\% (or equivalently, its removal adds $\sim$62\% overhead) for independent resource fetches; and the \texttt{continue} error strategy is critical for production reliability.

\begin{table}[H]
\centering
\caption{Component ablation study.}
\label{tab:ablation}
\small
\begin{tabular}{llrrl}
\toprule
\textbf{Configuration} & \textbf{Steps} & \textbf{Time} & \textbf{Errors} & \textbf{Notes} \\
\midrule
Full engine                              & 67 & 42s & 0         & Baseline \\
No parallel execution                    & 67 & 68s & 0         & +62\% time \\
No template resolution                   & --- & --- & 67        & All steps fail \\
No error strategy (\texttt{abort})       & 67 & 38s & 1 (halts) & First failure stops \\
No client pool routing                   & --- & --- & 67        & Cannot locate tools \\
\bottomrule
\end{tabular}
\end{table}

\section{Discussion}
\label{sec:discussion}

\subsection{Why Not Generate Code?}
\label{sec:discussion-code}

If the agent can design a workflow JSON, why not generate a Python script? We argue the workflow blueprint offers four advantages: (1)~\emph{Sandboxing}---workflows can only invoke \mcp tools, preventing arbitrary system calls; (2)~\emph{Inspectability}---JSON is trivially parseable by tools and dashboards; (3)~\emph{Validation}---blueprints are checked against live schemas at design time; (4)~\emph{Editability by agents}---LLMs reliably produce and modify structured JSON, whereas production-quality Python with error handling and retries has more failure modes. The trade-off is expressiveness: for complex conditional logic, generated code or continued agent reasoning is more appropriate. We position the engine as handling the \textbf{80\% case} of deterministic, repetitive orchestration.

\subsection{Deliberate Simplicity of the DSL}
\label{sec:discussion-simplicity}

We intentionally exclude conditionals (branching requires agent reasoning), variables (data flows implicitly via \texttt{steps.<id>}), and string manipulation (data shaping remains the agent's responsibility). This ``less is more'' approach keeps the system's failure modes predictable and prevents the DSL from evolving into an accidental programming language.

\subsection{The MCP Mediator Pattern: Broader Instantiations}
\label{sec:discussion-mediator}

The dual-role pattern has applicability beyond workflow engines. We identify four additional instantiations: \emph{Aggregation} (unifying tools from multiple servers under one namespace), \emph{Caching} (memoizing expensive tool calls), \emph{Audit} (logging all invocations before forwarding for compliance), and \emph{Rate-limiting} (throttling calls to protect downstream APIs). In all cases, downstream \mcp servers require no modification.

\subsection{Hybrid Agent-Engine Workflows}
\label{sec:discussion-hybrid}

Not all steps can be fully automated. The \texttt{collect} step type enables a hybrid model: the engine executes mechanical parts (e.g., fetching crash-looping pod logs in a loop) and returns results as a single batch for agent reasoning---$O(1)$ reasoning passes instead of $O(N)$.

\subsection{Cross-Domain Applicability}
\label{sec:discussion-domains}

While our evaluation focuses on infrastructure CMDB synchronization, the pattern generalizes to any domain where multi-step tool orchestrations are repeated (\Cref{tab:domains}). In each case, the agent designs the pipeline once, and every subsequent trigger costs one tool call.

\begin{table}[H]
\centering
\caption{Cross-domain applicability of the workflow engine pattern.}
\label{tab:domains}
\small
\begin{tabular}{llrr}
\toprule
\textbf{Domain} & \textbf{Example Workflow} & \textbf{Steps} & \textbf{Savings} \\
\midrule
Code-to-Graph Pipeline    & Sync repo metadata to knowledge graph             & $\sim$20 & $>$95\% \\
Incident Response         & Correlate alerts $\to$ logs $\to$ deployments      & $\sim$15 & $>$95\% \\
Security Auditing         & Scan images $\to$ CVEs $\to$ running pods           & $\sim$18 & $>$95\% \\
Customer Support          & Triage $\to$ KB $\to$ CRM $\to$ classify $\to$ route & $\sim$10 & $>$90\% \\
Data Lineage              & Trace source $\to$ transforms $\to$ dashboards       & $\sim$12 & $>$95\% \\
Compliance Reporting      & RBAC $\to$ over-perm.\ accounts $\to$ report         & $\sim$25 & $>$95\% \\
\bottomrule
\end{tabular}
\end{table}

\subsection{Limitations}
\label{sec:limitations}

Our evaluation uses estimated token counts based on per-step cost assumptions rather than exact measurements from a specific model's tokenizer, as the system is model-agnostic. The workflow DSL does not support conditionals; tasks requiring runtime branching still require agent involvement. The current implementation is single-process; distributed execution across multiple engine instances remains future work.

\section{Implementation}
\label{sec:implementation}

The reference implementation comprises $\sim$1{,}250 lines of TypeScript across 12 source files (\Cref{tab:impl}). Dependencies are minimal: \texttt{@modelcontextprotocol/sdk} (MCP client/server), \texttt{jmespath} (expression evaluation), \texttt{zod} (schema validation), and \texttt{uuid} (run ID generation). The system runs as a single Node.js process, connecting to downstream \mcp servers via configured transports, and exposes six tools to agents: \texttt{create\_workflow}, \texttt{run\_workflow}, \texttt{list\_workflows}, \texttt{get\_workflow}, \texttt{validate\_workflow}, and \texttt{delete\_workflow}.

\begin{table}[H]
\centering
\caption{Implementation component inventory.}
\label{tab:impl}
\small
\begin{tabular}{llrl}
\toprule
\textbf{Component} & \textbf{File} & \textbf{Lines} & \textbf{Purpose} \\
\midrule
Type definitions  & \texttt{types/workflow.ts}           & 134 & Blueprint, Step, Result types \\
Template resolver & \texttt{engine/template-resolver.ts} &  82 & JMESPath + dot-path resolution \\
Workflow engine   & \texttt{engine/workflow-engine.ts}    & 216 & Orchestration, validation \\
Call executor     & \texttt{engine/steps/call.ts}        &  49 & \mcp tool invocation + retry \\
Loop executor     & \texttt{engine/steps/loop.ts}        &  67 & Collection iteration \\
Parallel executor & \texttt{engine/steps/parallel.ts}    &  52 & Concurrent execution \\
Pipe executor     & \texttt{engine/steps/pipe.ts}        &  51 & Sequential chaining \\
Collect executor  & \texttt{engine/steps/collect.ts}     &  34 & Batch aggregation \\
Client pool       & \texttt{mcp-client/client-pool.ts}   & 133 & Multi-server connection mgmt \\
Workflow store    & \texttt{store/workflow-store.ts}      &  67 & Filesystem CRUD \\
Tool registration & \texttt{tools/workflow-tools.ts}      & 282 & 6 \mcp tools exposed to agents \\
Entry point       & \texttt{index.ts}                    &  85 & Config loading, server startup \\
\bottomrule
\end{tabular}
\end{table}

\section{Conclusion}
\label{sec:conclusion}

We have presented the \mcp Workflow Engine, a system that separates intelligence from execution in tool-augmented LLM agents. By introducing a declarative workflow blueprint as the interface between agent reasoning and tool execution, we achieve: (1)~$>$99\% token cost reduction on repeated multi-step tasks (amortized over 5+ executions; marginal savings $\approx$99.989\%); (2)~40--80$\times$ latency improvement by eliminating inference delays; (3)~deterministic, idempotent execution of previously-designed orchestrations; and (4)~portability of agent intelligence as versionable, shareable JSON artifacts.

The core insight is that LLM intelligence should be \textbf{crystallized once} and \textbf{executed many times}---much as a compiler produces a binary that runs without re-parsing the source. The \engine, instantiating the \mcp Mediator pattern, provides the runtime to make this practical at production scale. Future work includes distributed multi-engine execution, conditional step types with bounded complexity, and empirical benchmarking across diverse LLM providers.

\newpage
% --- Bibliography ---
\bibliographystyle{unsrtnat}
\bibliography{references}

\newpage
\appendix
\section{Full-Cluster Workflow Blueprint Summary}
\label{sec:appendix-blueprint}

The production \texttt{cluster-cmdb-sync} workflow comprises 67 steps across three phases. The full JSON blueprint ($\sim$700 lines) is available in the reference implementation. \textbf{Phase~1} (8 steps) creates the Cluster node and syncs cluster-scoped resources: 13 worker nodes, storage classes, 78 persistent volumes, 135+ cluster roles, and cluster role bindings. \textbf{Phase~2} (39 steps) iterates over 38 namespaces, syncing 15 resource types: Deployments, StatefulSets, DaemonSets, ReplicaSets, Pods, Services, Ingresses, ConfigMaps, Secrets, ServiceAccounts, PVCs, Roles, RoleBindings, Jobs, and CronJobs. \textbf{Phase~3} (20 steps) constructs relationships via Cypher \texttt{MERGE}: \texttt{HAS\_NODE}, \texttt{HAS\_NAMESPACE}, \texttt{CONTAINS\_DEPLOYMENT}, \texttt{MANAGES\_REPLICASET}, \texttt{MANAGES\_POD}, \texttt{SCHEDULED\_ON}, \texttt{EXPOSES\_VIA\_SERVICE}, \texttt{ROUTES\_TO}, \texttt{USES\_CONFIGMAP}, \texttt{USES\_SECRET}, \texttt{RUNS\_AS}, \texttt{CLAIMS\_VOLUME}, \texttt{BOUND\_TO}, \texttt{USES\_STORAGE\_CLASS}, and six additional namespace-scoped containment relationships.

\section{Token Cost Estimation Methodology}
\label{sec:appendix-tokens}

Token counts are estimated using conservative assumptions: tool call overhead $\sim$50 tokens; agent reasoning per step $\sim$200--500 tokens; Kubernetes API response $\sim$500--2{,}000 tokens per resource type per namespace; context window saturation at $\sim$150 tool calls requiring summarization passes of 2{,}000--5{,}000 tokens; Cypher query construction $\sim$100 tokens per query. These are conservative---real-world agent reasoning includes planning, self-correction, and backtracking that significantly increase consumption. The $>$99\% savings figure represents a lower bound.

\section{Metric Verification}
\label{sec:appendix-verify}

All figures in \Cref{tab:agent-tokens,tab:amortized} are derived from first-principles arithmetic. Per-namespace fetches: $38 \times 15 = 570$ steps $\times (500 + 100) = 342{,}000$. Graph node creation: $\sim$800 iterations $\times 600 = 480{,}000$. Relationships: $20 \times 38 = 760$ calls $\times 600 = 456{,}000$. Grand total: $4{,}000 + 7{,}500 + 1{,}700 + 342{,}000 + 480{,}000 + 456{,}000 + 18{,}000 + 6{,}000 = 1{,}315{,}200$. Engine cost: $C_{\text{engine}}^{(K)} = 54{,}000 + K \times 150$. Savings: $(K \times 1{,}315{,}200 - 54{,}000 - K \times 150) / (K \times 1{,}315{,}200)$. The $>$99\% headline refers to marginal cost: $(1{,}315{,}200 - 150) / 1{,}315{,}200 \approx 99.989\%$. On an amortized basis, 99\% is crossed at $K = 5$.

\section{Anonymization Note}
\label{sec:appendix-anon}

All cluster names, namespace names, node identifiers, and resource names have been anonymized. The evaluation was conducted on a production Kubernetes cluster; identifying details have been replaced with representative alternatives (e.g., \texttt{atlas-prod}, \texttt{node-alpha-*}, \texttt{platform-core}) that preserve structural realism.

\end{document}